\documentclass[5p]{elsarticle}

\usepackage{graphics}
\usepackage{amssymb}

\begin{document}

\begin{frontmatter}

\title{Properties of the half-filled Hubbard model investigated by the strong coupling diagram technique}

\author{A. Sherman}
\ead{alexei@fi.tartu.ee}
\address{Institute of Physics, University of Tartu, Ravila 14c,
50411 Tartu, Estonia}

\begin{abstract}
The equation for the electron Green's function of the fermionic Hubbard model, derived using the strong coupling diagram technique, is solved self-consistently for the near-neighbor form of the kinetic energy and for half-filling. In this case the Mott transition occurs at the Hubbard repulsion $U_c\approx 6.96t$, where $t$ is the hopping constant. The calculated spectral functions, density of states and momentum distribution are compared with results of Monte Carlo simulations. A satisfactory agreement was found for $U>U_c$ and for temperatures, at which magnetic ordering and spin correlations are suppressed. For $U<U_c$ and lower temperatures the theory describes qualitatively correctly positions and widths of spectral continua, variations of spectral shapes and occupation numbers with changing wave vector and repulsion. The locations of spectral maxima turn out to be close to the positions of $\delta$-function peaks in the Hubbard-I approximation.
\end{abstract}


\end{frontmatter}

\section{Introduction}
The repulsive fermionic Hubbard model is one of the main models describing strong electron correlations in crystals. The model and its generalizations are frequently used for interpreting experimental results in cuprate perovskites and other compounds with strong correlations. The strong coupling diagram technique \cite{Vladimir,Metzner,Craco,Pairault,Sherman06} is one of the approximate approaches used for investigating the model. The method is based on the serial expansion in powers of the electron kinetic energy. The elements of the arising diagram technique are on-site cumulants of electron creation and annihilation operators and hopping lines connecting cumulants on different sites. As in the diagram technique with the expansion in powers of an interaction, in the considered approach the linked-cluster theorem allows one to discard disconnected diagrams and to carry out partial summations in remaining connected diagrams. As a consequence the electron Green's function is expressed in the form of the Larkin equation containing the initial electron dispersion and the irreducible part -- the sum of all irreducible diagrams without external ends. In spite of the fact that by the construction the approach is intended for the case of strong coupling, when the Hubbard repulsion $U$ is approximately equal to or larger than the initial bandwidth $\Delta$, it gives the correct result in the limit $U\rightarrow 0$. Hence the approach provides an interpolation between the limits of weak and strong correlations.

Using this method, in recent work \cite{Sherman15} the equation for the electron Green's function was obtained by keeping terms of the lowest two orders in the irreducible part of the Larkin equation. Self-consistent calculations performed for the semi-elliptical initial band showed that at half-filling the approximation describes the Mott transition, which occurs at $U_c=\sqrt{3}\Delta/2$. This value coincides with the critical repulsion obtained for the same initial band in the Hubbard-III approximation \cite{Hubbard64}. Supplementary solutions of the equation and the evolution of the density of states (DOS) with deviations from half-filling were also investigated in \cite{Sherman15}.

In the present work the equation of Ref.~\cite{Sherman15} is self-consistently solved for the near-neigh\-bor form of the kinetic energy and for half-filling. In this case one can calculate not only the DOS, as for the semi-elliptical band, but also spectral functions and related quantities, which furnish an opportunity to compare obtained results with data of Monte Carlo simulations. For this initial band the Mott transition was found to occur at $U_c\approx 6.96t$, where $t$ is the hopping constant. The ratio of this critical repulsion to the initial bandwidth $\Delta=8t$ appears to be very close to that found in \cite{Sherman15} for the semi-elliptical initial band. For the Hubbard repulsions $U>U_c$ the calculated spectral functions are in satisfactory agreement with Monte Carlo spectra obtained for temperatures $T\gtrsim t$, for which the magnetic ordering and spin correlations are suppressed. For smaller values of $U$ and $T$ the agreement deteriorates. However, even in this case the theory describes qualitatively correctly the location of spectral continua, shapes of spectra and their variation with changing wave vector and repulsion. The degradation of the agreement with the numerical experiments for small $U$ and $T$ can be expected because the irreducible part is local in the used approximation. This means it does not take into account the interaction of electrons with magnetic ordering, spin and charge fluctuations. The location of maxima in the calculated spectra turn out to be close to the positions of the $\delta$-function peaks of the Hubbard-I approximation \cite{Hubbard63}. One can say that the considered approach is an improvement of this latter approximation, which introduces finite widths into the peaks, describes the Mott gap arising in the continuous spectrum at the Fermi level when $U$ exceeds $U_c$, and reduces to the uncorrelated solution when $U\rightarrow 0$. Besides the spectral functions, the DOS and the momentum distribution were calculated for several values of $U$. These dependencies reproduce qualitatively Monte Carlo data also. In the range $0<U<U_c$ the obtained solution is not a Fermi liquid, since the imaginary part of the self-energy ${\rm Im\,\Sigma}$ is nonzero at the Fermi level, though it is small for $U\ll U_c$ and the frequency dependence of ${\rm Im\,\Sigma}$ may resemble the quadratic dependence of the weak-coupling case. As $U$ approaches the critical value from below, ${\rm Im\,\Sigma}$ diverges as $(U_c-U)^{-0.75}$.

\section{Main formulas}\label{sec2}
In this section some equations derived in Ref.~\cite{Sherman15} are reproduced and converted to the form, which is convenient for calculations. The electron Green's function
\begin{equation}\label{Green}
G({\bf n'\tau',l\tau})=\langle{\cal T}\bar{a}_{\bf n'\sigma}(\tau')
a_{\bf n\sigma}(\tau)\rangle
\end{equation}
is considered, where the angular brackets denote the statistical averaging with the Hamiltonian
\begin{eqnarray}\label{Hamiltonian}
H&=&\sum_{\bf nn'\sigma}t_{\bf nn'}a^\dagger_{\bf n\sigma}a_{\bf n'\sigma} +\frac{U}{2}\sum_{\bf n\sigma}n_{\bf n\sigma}n_{\bf n,-\sigma}\nonumber\\
&&-\mu\sum_{\bf n\sigma}n_{\bf n\sigma},
\end{eqnarray}
$t_{\bf nn'}$ is the hopping constants, the operator $a^\dagger_{\bf n\sigma}$ creates an electron on the site {\bf n} of the two-dimensional (2D) square lattice with the spin projection $\sigma=\pm 1$, the electron number operator $n_{\bf n\sigma}=a^\dagger_{\bf n\sigma}a_{\bf n\sigma}$, $\mu$ is the chemical potential, ${\cal T}$ is the time-ordering operator which
arranges operators from right to left in ascending order of times $\tau$,
$a_{\bf n\sigma}(\tau)=\exp(H\tau)a_{\bf n\sigma}\exp(-H\tau)$ and $\bar{a}_{\bf n\sigma}(\tau)=\exp(H\tau)a^\dagger_{\bf n\sigma}\exp(-H\tau)$. Green's function (\ref{Green}) does not depend on the spin projection, and it was omitted in the function notation.

In the strong coupling diagram technique, after the summation of all diagrams the Fourier transform of Green's function (\ref{Green}) acquires the form of the Larkin equation
\begin{equation}\label{Larkin}
G({\bf k},i\omega_l)=\frac{K({\bf k},i\omega_l)}{1-t_{\bf k}K({\bf k},i\omega_l)},
\end{equation}
where ${\bf k}$ is the 2D wave vector, $\omega_l=(2l+1)\pi T$ is the Matsubara frequency, $t_{\bf k}=\sum_{\bf n}\exp[i{\bf k(n-n')}]t_{\bf nn'}$ and $K({\bf k},i\omega_l)$ is the irreducible part -- the sum of all irreducible diagrams without external ends. A diagram is said to be an irreducible one if it cannot be divided into two disconnected parts by cutting some hopping line $t_{\bf nn'}$.

In Ref.~\cite{Sherman15} and in this work $K({\bf k},i\omega_l)$ is approximated by the sum of two terms of the lowest orders in powers of $t_{\bf k}$. These terms contain cumulants of the first and second orders. The second term of the approximate $K({\bf k},i\omega_l)$, which contains the second-order cumulant, includes also a hopping-line loop. Using the possibility of the partial summation of diagrams we transform this bare hopping line into the dressed one,
\begin{eqnarray}\label{hopping}
&&\theta({\bf n}\tau,{\bf n'}\tau')=t_{\bf nn'}\delta(\tau-\tau')+\sum_{\bf mm'}t_{\bf nm}\nonumber\\
&&\quad\quad\quad\times\int_0^\beta d\upsilon\, K({\bf m}\tau,{\bf m'}\upsilon)\, \theta({\bf m'}\upsilon,{\bf n'}\tau'),
\end{eqnarray}
where $\beta=1/T$. With this substitution, for a chemical potential in the range
\begin{equation}\label{mu_range}
\mu\gg T,\quad U-\mu\gg T
\end{equation}
the expression for $K({\bf k},i\omega_l)$ reads
\begin{eqnarray}\label{K_gen}
&&K(i\omega_l)=\frac{1}{2}\left[g_{01}(i\omega_l)+g_{12}(i\omega_l)\right]\nonumber\\
&&\quad+\frac{3}{4}F^2(i\omega_l)\phi(i\omega_l) -\frac{s_1}{2}F(i\omega_l)-\frac{s_2}{2}J(i\omega_l),
\end{eqnarray}
where
\begin{eqnarray}\label{variables}
&&g_{01}(i\omega_l)=\left(i\omega_l+\mu\right)^{-1},\nonumber\\ &&g_{12}(i\omega_l)=\left(i\omega_l+\mu-U\right)^{-1},\nonumber\\ &&F(i\omega_l)=g_{01}(i\omega_l)-g_{12}(i\omega_l),\nonumber\\
&&J(i\omega_l)=g^2_{01}(i\omega_l)-F(i\omega_l)g_{12}(i\omega_l),\\
&&\phi(i\omega_l)=N^{-1}\sum_{\bf k}t^2_{\bf k}G({\bf k},i\omega_l),\nonumber\\
&&s_1=T\sum_l J(i\omega_l)\phi(i\omega_l),\nonumber\\
&&s_2=T\sum_l F(i\omega_l)\phi(i\omega_l)\nonumber
\end{eqnarray}
and $N$ is the number of sites. We chose $t_{\bf nn}=0$ and used this relation in deriving Eq.~(\ref{K_gen}). Equations (\ref{Larkin}), (\ref{K_gen}) and (\ref{variables}) form the closed set of equation for calculating Green's function (\ref{Green}). Notice that irreducible part (\ref{K_gen}) does not depend on momentum and, therefore, the approximation does not take into account the interactions of electrons with the magnetic ordering, spin and charge fluctuations. These interactions are described by the sum of diagrams containing ladders, which correspond to the dynamic spin and charge susceptibilities \cite{Sherman07}.

In this work we consider the case of half-filling, when $\mu=u\equiv U/2$. Besides, it is supposed that the hopping constants $t_{\bf nn'}$ are nonzero only for nearest neighbor sites, which results in $t_{\bf k}=2t\left[\cos(k_x a)+\cos(k_y a)\right]$, where $a$ is an intersite distance. In the following, we set $t$ and $a$ as the units of energy and length. In the considered case the above formulas are essentially simplified, since the sums $s_1$ and $s_2$, Eq.~(\ref{variables}), vanish due to the fact that $F(i\omega_l)$ and $J(i\omega_l)$ in them are even functions of $\omega_l$, while $\phi(i\omega_l)$ is an odd one. After the analytic continuation to real frequencies $\omega$ the equation for $K(\omega)$ can be written as
\begin{eqnarray}\label{K_hf}
K(\omega)&=&\frac{\omega}{\omega^2-u^2}\nonumber\\
&&+\frac{3u^2}{N(\omega^2-u^2)^2}\sum_{\bf k}\frac{t_{\bf k}}{1-t_{\bf k}K(\omega)}.
\end{eqnarray}
Notice that in the used approximation $K(\omega)$ does not depend on temperature.

From Eq.~(\ref{Larkin}) one can see that $K(\omega)$ coincides with Green's function for wave vectors satisfying the condition $t_{\bf k}=0$. For the considered initial band these momenta form line segments connecting centers of sides of the Brillouin zone $(\pm\pi,0)$ and $(0,\pm\pi)$. Here we face problem because $K(\omega)$ is not an analytic function in the upper complex frequency half-plane, as it must be for a retarded Greens function. Indeed, let us consider an auxiliary function $\kappa(\omega)$, which is connected with the irreducible part by the relations
\begin{eqnarray}
&&{\rm Re}\,\kappa(\omega)=(\omega^2-u^2)\,{\rm Re}\,K(\omega),\label{rekappa}\\
&&{\rm Im}\,\kappa(\omega)=|\omega^2-u^2|\,{\rm Im}\,K(\omega).\label{imkappa}
\end{eqnarray}
The equation for $\kappa(\omega)$ reads
\begin{equation}\label{eq_kpp}
\kappa(\omega)=\omega+\frac{3u^2}{N}\sum_{\bf k}\frac{t_{\bf k}}{\omega^2-u^2-t_{\bf k} \kappa(\omega)}.
\end{equation}
This equation can be solved analytically in the points $\omega=\pm u$ with the result
$$\kappa(\pm\omega)=\left(\pm1-i\sqrt{11}\right)\frac{u}{2},$$
where the sign of ${\rm Im}\,\kappa(\pm u)$ was chosen from the requirement that ${\rm Im}\,\kappa(\omega)$, like ${\rm Im}\, K(\omega)$, be negative on the real axis. The quantity $\kappa(\omega)$ is a continuous function near the points $\omega=\pm u$. As a consequence ${\rm Im}\, K(\omega)$ behaves like $-(\sqrt{11}u/2)|\omega^2-u^2|^{-1}$ near these points, and the integral
$${\cal I}={\cal P}\int_{-\infty}^\infty\frac{d\omega'}{\pi}\frac{{\rm Im}\,K(\omega')}{\omega'-\omega}$$
in the Kramers-Kronig relations diverges. Here ${\cal P}$ means the Cauchy principal value of the integral, which is related to the singularity at $\omega'=\omega$. Due to the divergence of ${\cal I}$ the Kramers-Kronig relations do not fulfil and $K(\omega)$ is not an analytic function in the upper half-plane.

The simplest way to overcome this difficulty is to introduce an artificial broadening $\eta$ by substituting relation (\ref{imkappa}) by the equation
\begin{equation}\label{broadening}
{\rm Im}\,K(\omega)=\frac{{\rm Im}\,\kappa(\omega)}{\sqrt{(\omega^2-u^2)^2+\eta^4}}.
\end{equation}
With function (\ref{broadening}) the integral ${\cal I}$ becomes convergent, and the Kramers-Kronig relations can be used for calculating ${\rm Re}\, K(\omega)$ from this ${\rm Im}\, K(\omega)$, which assures the analyticity of $K(\omega)$ in the upper half-plane. Before this step ${\rm Im}\, K(\omega)$ has to be normalized for a given $\eta$,
\begin{equation}\label{normalization}
\int_{-\infty}^\infty d\omega\,{\rm Im}\, K(\omega)=-\pi.
\end{equation}
This ensures the normalization of the spectral function $A({\bf k}\omega)=-{\rm Im}\,G({\bf k}\omega)/\pi$ to unity that for wave vectors satisfying the condition $t_{\bf k}=0$, when $G({\bf k}\omega)=K(\omega)$. We proved that the fulfillment of condition (\ref{normalization}) assures the normalization of spectral functions for all points in the Brillouin zone with good accuracy.

\begin{figure}
\centerline{\resizebox{0.95\columnwidth}{!}{\includegraphics{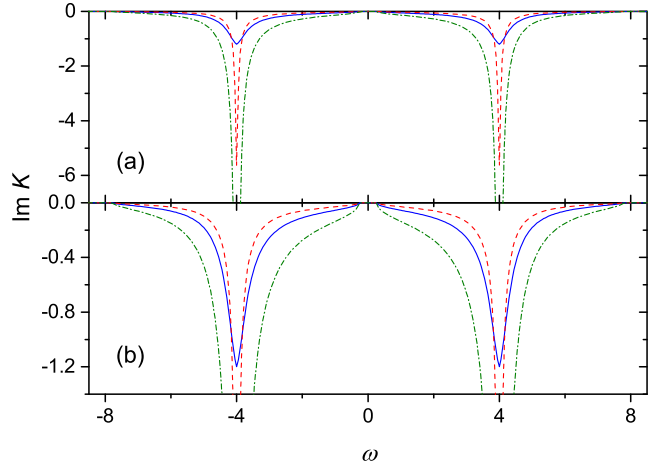}}}
\caption{The imaginary part of $K(\omega)$ calculated from Eq.~(\protect\ref{K_hf}) (olive dash-dotted line) and from Eqs.~(\protect\ref{eq_kpp}) and (\protect\ref{broadening}) for $\eta=0.5$ (red dashed line) and $\eta=1.5$ (blue solid line). The two latter curves were normalized in accord with Eq.~(\protect\ref{normalization}). $U=8$. The part (b) of the figure shows curves in part (a) in a larger $y$ scale.} \label{Fig1}
\end{figure}
The imaginary parts of $K(\omega)$ obtained from Eq.~(\ref{K_hf}) and from Eqs.~(\ref{eq_kpp}), (\ref{broadening}) and (\ref{normalization}) for different $\eta$ are compared in Fig.~\ref{Fig1}. Positions and widths of continua, the existence and magnitude of the Mott gap in the spectrum are determined by the quantity $\kappa(\omega)$ in Eq.~(\ref{eq_kpp}). The substitution of Eq.~(\ref{imkappa}) with (\ref{broadening}) cannot change them. As seen in the figure, the introduction of $\eta$ restricts intensities of maxima and determines their widths. However, their frequencies do not change with $\eta$. Notice that for wave vectors, for which $t_{\bf k}\neq 0$, the frequencies of the spectral maxima are slightly changed with $\eta$.

Green's functions and related quantities of the next section were obtained from the self-consistent solutions of Eq.~(\ref{eq_kpp}). In this calculations the Newton-Raphson method or direct iteration were used. For a selected broadening $\eta$, which usually varied from several tenth to several units, ${\rm Im}\,K(\omega)$ was calculated from Eq.~(\ref{broadening}), and after its normalization (\ref{normalization}) the real part of $K(\omega)$ was found from the Kramers-Kronig relations. Green's function follows from Eq.~(\ref{Larkin}).

\section{Results and discussion}
\subsection{The Mott transition}
For the considered model the gap in the dependence ${\rm Im}\,\kappa(\omega)$ appears at the Fermi level when the Hubbard repulsion exceeds the value $U_c\approx 7$. This gap reveals itself in ${\rm Im}\,K(\omega)$ (see, e.g., Fig.~\ref{Fig1}), in spectral functions and in the DOS. Since Eq.~(\ref{eq_kpp}) does not contain $\eta$, the critical value of the repulsion and the gap magnitude do not depend on this parameter. The imaginary part of the self-energy at the Fermi level diverges as $U$ approaches $U_c$ (see Subsection~\ref{self-energy}). This fact allows us to define $U_c$ more exactly: $U_c\approx 6.96$. The ratio of this parameter to the bandwidth $\Delta=8$ is close to the analogous ratio $U_c/\Delta=\sqrt{3}/2$ obtained in the Hubbard-III approximation \cite{Hubbard64} and in Ref.~\cite{Sherman15} for the semi-elliptical DOS.

\begin{figure}
\centerline{\resizebox{0.75\columnwidth}{!}{\includegraphics{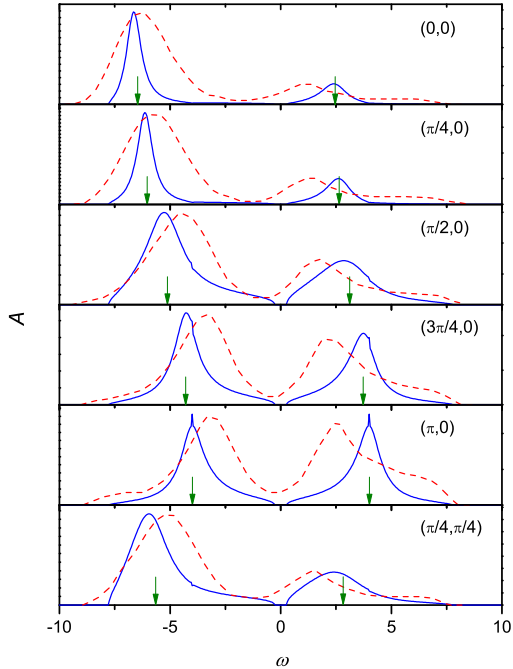}}}
\caption{The spectral function $A({\bf k}\omega)$ calculated in a $8\times 8$ lattice for wave vectors shown in the panels and $U=8$ (blue solid lines). Arrows indicate the positions of $\delta$-function peaks in the Hubbard-I approximation. Red dashed lines show results of Monte Carlo simulations performed in Ref.~\cite{Groeber} for the same lattice and $U$, and for $T=1$.} \label{Fig2}
\end{figure}

\begin{figure}
\centerline{\resizebox{0.765\columnwidth}{!}{\includegraphics{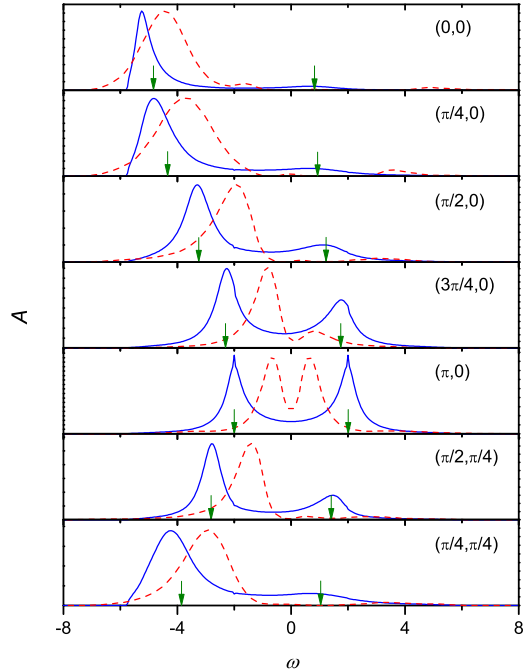}}}
\caption{The spectral function $A({\bf k}\omega)$ calculated in a $8\times 8$ lattice for wave vectors shown in the panels and $U=4$ (blue solid lines). Arrows indicate the positions of $\delta$-function peaks in the Hubbard-I approximation. Red dashed lines show results of Monte Carlo simulations performed in Ref.~\cite{Moukouri} for the same lattice and $U$, and for $T=0.2$.} \label{Fig3}
\end{figure}

\subsection{Spectral functions}
The spectral functions calculated in a $8\times 8$ lattice for two values of the repulsion, $U=8>U_c$ and $U=4<U_c$, are shown in Figs.~\ref{Fig2} and \ref{Fig3}. For this lattice, spectra for nearly all independent wave vectors are shown in the figures. The spectrum for ${\bf k}=(\pi/2,\pi/4)$ and $U=8$ is not shown because we could not find Monte Carlo data for these parameters. Spectra for only one momentum on the boundary of the magnetic Brillouin zone, ${\bf k}=(\pi,0)$, are depicted in both figure, since, as mentioned above, for all these momenta the spectra are identical and coincides with $-{\rm Im}K(\omega)/\pi$. In Monte Carlo simulations \cite{Groeber,Moukouri}, these spectra are closely similar to each other also. Spectra in other points of the Brillouin zone are either identical to those shown in the pictures or can be obtained from them by the specular reflection in the $\omega=0$ axis.

In these figures arrows indicate positions of $\delta$-function peaks of the Hubbard-I approximation,
$$\varepsilon_{\bf k}=\frac{1}{2}\left(t_{\bf k}\pm\sqrt{t_{\bf k}^2+U^2}\right).$$
As one can see, our calculated spectral maxima are close to these frequencies. Hence, the used approach can be considered as an improvement of the Hubbard-I approximation, which give maxima at nearly the same locations, and additionally it introduces finite widths of these maxima and describes the Mott gap arising in the continuous spectrum at the Fermi level when $U$ exceeds $U_c$.

In the figures, the calculated spectra are compared with the results of Monte Carlo simulations performed in Refs.~\cite{Groeber,Moukouri}. As seen from Fig.~\ref{Fig2}, the agreement is wholly satisfactory in the case of $U>U_c$ and high temperatures, when magnetic ordering and spin correlations are suppressed. The ability of the Hubbard-I approximation with artificially broadened peaks to reproduce maxima in Monte Carlo spectra in these conditions was also pointed out in Ref.~\cite{Groeber}. Notice, however, that the used approach reproduce satisfactorily not only the location of maxima but also the width and location of spectral bands. The agreement may be even better than that shown in Fig.~\ref{Fig2} because Monte Carlo spectra are somewhat distorted by errors, which among others are introduced by a maximum-entropy analytic continuation procedure. In particular, we suppose that the spectrum at ${\bf k}=(\pi,0)$ is symmetric with respect to $\omega=0$, like our spectrum for this momentum. For the $(\pi/2,\pi/2)$ spectrum, which is similar to the $(\pi,0)$ spectrum and also slightly asymmetric in the Monte Carlo results of Ref.~\cite{Groeber}, this fact follows from the particle-hole symmetry of the problem. Besides, in Monte Carlo data of Refs.~\cite{Moukouri,Bulut} the $(\pi,0)$ spectrum is symmetric with respect to $\omega=0$ (cf.\ Fig.~\ref{Fig3}).

With decreasing temperature intensities in Monte Carlo spectra are redistributed and an additional structure appears. As a consequence the agreement of our results with these spectra is impaired. This fact is apparently connected with an increased influence of the magnetic ordering and spin correlations, which start to grow with lowering $T$. As mentioned above, interactions of electrons with the magnetic ordering and spin fluctuations are not taken into account by approximations with a local irreducible part or self-energy. As the repulsion is decreased below $U_c$ additionally charge fluctuations, which are also ignored in the considered approach, become more pronounced. As a result, the agreement deteriorates even more, as seen in Fig.~\ref{Fig3}. However, even in this, the worst for the used approximation, case the theory describes qualitatively correctly the location of the spectral continua, general spectral shape and the variation of maxima with wave vector and $U$. Notice also that in the case $U\ll U_c$ and low temperatures the nesting of the Fermi surface in the considered model will lead to the appearance of the antiferromagnetic gap on the Fermi level in the electron spectrum \cite{Penn}. This process is not described by the used approximation.

Sharp feature near $\omega=\pm U/2$, which are visible in some spectra in the figures, are connected with the smallness of the used lattice. They disappear in larger lattices [cf.\ the $(\pi,0)$ spectrum in Fig.~\ref{Fig2} with Fig.~\ref{Fig1}, which was calculated in a $240\times240$ lattice]. In other respects spectra for larger lattices are close to those shown in Figs.~\ref{Fig2} and \ref{Fig3} for the same momenta.  Spectra cease to change perceptibly for lattices larger than $32\times32$ sites.

\subsection{The density of states}
\begin{figure}
\centerline{\resizebox{0.7\columnwidth}{!}{\includegraphics{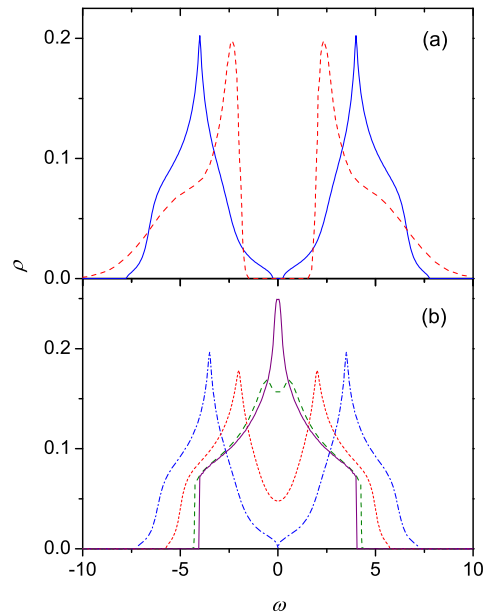}}}
\caption{(a) The density of states $\rho(\omega)$ calculated in a $12\times 12$ lattice for $U=8$ (blue solid line) in comparison with the Monte Carlo result of Ref.~\cite{Bulut}, obtained for the same lattice and repulsion, and for $T=0.125$ (red dashed line). (b) Densities of states calculated for $U=7$ (blue dash-dotted line), 4 (red short-dashed line), 1 (olive dashed line) and 0.1 (purple solid line) in a $400\times400$ lattice.} \label{Fig4}
\end{figure}

\begin{figure}
\centerline{\resizebox{0.8\columnwidth}{!}{\includegraphics{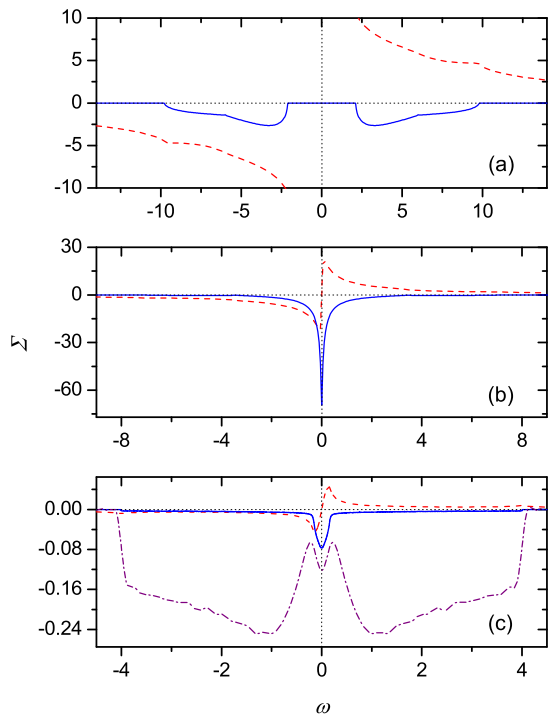}}}
\caption{Real (dashed red lines) and imaginary (solid blue lines) parts of the self-energy for $U=12$ and $\eta=3$ (a), $U=6.8$ and $\eta=0.5$ (b), $U=0.1$ and $\eta=0.1$ (c). In the latter panel ${\rm Im}\Sigma(\omega)$ for $U=0.1$ and $\eta=1$ is also shown by the purple dash-dotted line. Calculations were carried out in a $400\times400$ lattice.} \label{Fig5}
\end{figure}

In Fig.~\ref{Fig4}(a) the DOS $\rho(\omega)=N^{-1}\sum_{\bf k}A({\bf k}\omega)$, calculated in a $12\times 12$ lattice for $U=8$, is compared with the Monte Carlo result of Ref.~\cite{Bulut}, which was obtained for the same lattice and repulsion. The curves are similar, though the locations of maxima are somewhat shifted. As in the previous section, the difference is apparently connected with comparatively low temperature of the Monte Carlo simulations and the related perceptible spin correlations.

The evolution of the DOS with changing $U$ is shown in Fig.~\ref{Fig4}(b). As can be seen, the Mott gap disappears at $U\approx 7$. For smaller values of the repulsion only a dip at the Fermi level is seen in the DOS. For even smaller $U$ the dip is gradually transformed to the Van Hove maximum, and the DOS acquires the shape of the uncorrelated band. Except for the Van Hove singularity, the evolution resembles that obtained for the semi-elliptical initial band in Ref.~\cite{Sherman15}.

\subsection{The self-energy}\label{self-energy}
As follows from Eq.~(\ref{Larkin}), the self-energy is connected with the irreducible part by the relation $\Sigma({\bf k}\omega)=\omega-K^{-1}({\bf k}\omega)$ [for convenience here the chemical potential is included in the real part of $\Sigma({\bf k}\omega)$]. The real and imaginary parts of this quantity are shown in Fig.~\ref{Fig5}. In the case $U>U_c$, Fig.~\ref{Fig5}(a) the Mott gap is seen in the imaginary part and the real part diverges as $1/\omega$ when $\omega \rightarrow 0$. In its turn at $\omega=0$ the imaginary part diverges as $U\rightarrow U_c$ from below. As a demonstration of this fact an intensive peak of ${\rm Im}\Sigma(\omega)$ is shown in Fig.~\ref{Fig5}(b) for the value of the repulsion, which is slightly smaller than $U_c$. For small $U$ the shape of the imaginary part of the self-energy varies essentially with $\eta$. This is demonstrated in Fig.~\ref{Fig5}(c) for two values of $\eta$. Notice that for $\eta=1$ the electron damping has a dip with a small maximum on the bottom near the Fermi level. In the vicinity of the maximum the damping grows superlinearly, which resembles its quadratic growth with distance from the Fermi level in the case of weak correlations.

\begin{figure}
\centerline{\resizebox{0.8\columnwidth}{!}{\includegraphics{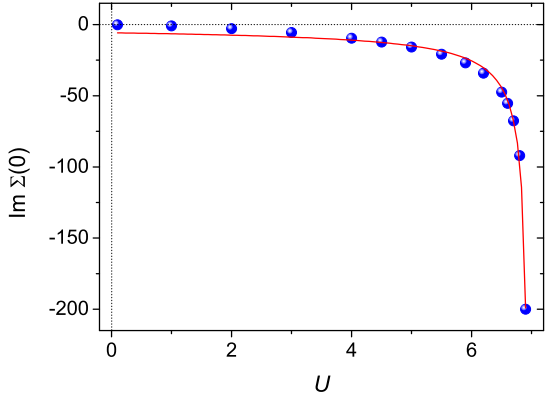}}}
\caption{The imaginary part of the self-energy at the Fermi level as a function of the repulsion for $\eta=0.3$ (symbols). The red curve shows fitting function (\protect\ref{divergence}) with $c=-24.78$, $U_c=6.96$ and $\nu=0.75$.} \label{Fig6}
\end{figure}
Figure~\ref{Fig6} demonstrates the behavior of the imaginary part of the self-energy at the Fermi level with changing $U$. As already mentioned, ${\rm Im}\Sigma(0)$ diverges as the repulsion approaches $U_c$ from below. Notice, however, that the value of ${\rm Im}\Sigma(0)$ varies with $\eta$ [see, e.g., Fig.~\ref{Fig5}(c)]. Therefore, the considered dependence was investigated for  several values of $\eta$. Near $U_c$ the dependence can be described by the relation
\begin{equation}\label{divergence}
{\rm Im}\Sigma(0)=\frac{c}{\left(U_c-U\right)^\nu}
\end{equation}
for all considered values of $\eta$. An example of a fitting of calculated data with this function is shown in Fig.~\ref{Fig6}. The parameter $c$ depends heavily on $\eta$, while the parameters $U_c$ and $\nu$ are nearly independent of it in the considered range $0.1\leq\eta\leq 0.9$ and equal to 6.96 and 0.75, respectively. This allows us to determine more exactly the value of the critical repulsion. The same dependence (\ref{divergence}) describes the behavior of ${\rm Im}\Sigma(0)$ in the case of the semi-infinite initial band, however, with the smaller exponent $\nu=0.5$ \cite{Gros,Sherman15}.

As for this latter band, ${\rm Im}\Sigma(0)$ is nonzero for $U$ in the range $0<U<U_c$, though it is small for small repulsions (see Fig.~\ref{Fig6}). Thus, the obtained solution is not a Fermi liquid. The fact that ${\rm Im}\Sigma(0)$ is nonzero for $U\ll U_c$ demonstrates that the used approximation does not describe the antiferromagnetic gap, which has to arise on the Fermi level due to the nested Fermi surface.

\subsection{The momentum distribution}
\begin{figure}
\centerline{\resizebox{0.75\columnwidth}{!}{\includegraphics{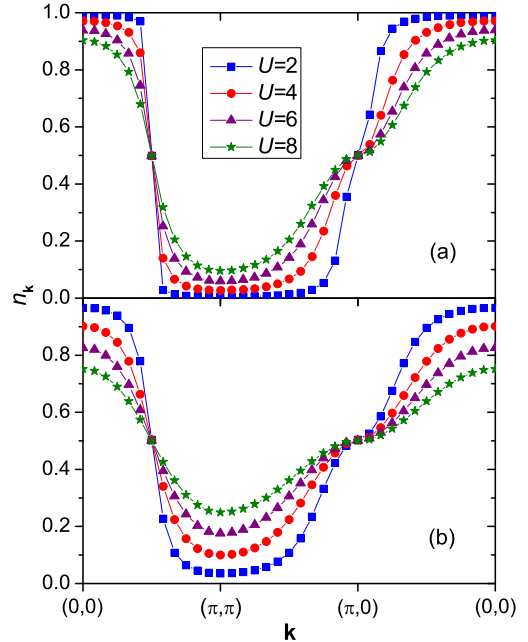}}}
\caption{The momentum distribution for wave vectors lying on the symmetry lines and for $U=2$ (blue square), 4 (red circles), 6 (purple triangles) and 8 (olive stars). (a) Monte Carlo simulations for $T$ ranging from 1/32 to 1/20 \protect\cite{Varney}, (b) results obtained from equations of Sec.~\protect\ref{sec2}.} \label{Fig7}
\end{figure}
In Fig.~\ref{Fig7} the momentum distribution
\begin{equation}\label{distribution}
n_{\bf k}=\left\langle a^\dagger_{\bf k\sigma}a_{\bf k\sigma}\right\rangle=\int_{-\infty}^\infty d\omega\frac{A({\bf k}\omega)}{\exp(\beta\omega)+1},
\end{equation}
calculated from equations of Sec.~\ref{sec2}, is compared with results of Monte Carlo simulation from Ref.~\cite{Varney}. Since in the considered approximation spectral functions do not depend on temperature, it was set to zero in Eq.~(\ref{distribution}). Both Monte Carlo and our calculations were carried out in a $24\times24$ lattice. As seen from Fig.~\ref{Fig7}, the theory reproduces correctly the variation of $n_{\bf k}$ with changing wave vector and repulsion. Dependencies shown in the figure demonstrate gradual blurring of the Fermi surface with increasing $U$ -- the growing repulsion leads to an increasing occupancy in the region of the Brillouin zone, which was empty at $U=0$. Apparently the observed differences between our results and Monte Carlo data are again connected with the fact that the considered approach neglects interactions with the magnetic ordering, spin and charge fluctuations, which have to be strong at low temperatures used in the Monte Carlo simulations.

\section{Summary}
In the present work, the strong coupling diagram technique was used for calculating Green's functions and related quantities of the two-dimensional fermionic Hubbard model. The case of half-filling and the near-neighbor form of the kinetic energy were considered. In this approach Green's function is described by the Larkin equation. In this work the irreducible part of this equation was approximated by terms of the two lowest orders in the serial expansion in powers of the hopping constant $t$. Using the possibility of the partial summation in the diagram technique the bare hopping line in the second term was substituted by the dressed one. As a consequence the Larkin equation becomes the closed relation for calculating Green's function. In this equation, the irreducible part is local, and, therefore, it neglects interactions of electrons with magnetic ordering, spin and charge fluctuations. The derived irreducible part has poles on the real frequency axis, which violate analytic properties of the retarded Green's function. To overcome this difficulty an artificial broadening and Kramers-Kronig relations were used. It was shown that this procedure influence only widths of spectral maxima, leaving the position of maxima, widths and locations of gaps and bands in the continuous spectrum intact.

It was found that the used approach describes the Mott transition, which occurs at the value of the Hubbard repulsion $U_c=6.96t$. The ration of $U_c$ to the bandwidth $8t$ is close to $\sqrt{3}/2$, the value obtained in the Hubbard-III approximation and in the same approach for the semi-elliptical initial band. Calculated locations of maxima in spectral functions appear to be close to those in the Hubbard-I approximation. Thus, the used approach can be considered as a development of this approximation, which corrects two its flows, transforming $\delta$-functions into spectral bands and describing the Mott transition. Besides, the approach gives the correct result in the limit of zero repulsion, thereby providing a way for interpolating between cases of week and strong correlations.

Comparison of our calculated spectral functions, density of states and momentum distribution with results of Monte Carlo simulations shows a satisfactory agreement in the cases $U>U_c$ and temperatures $T\gtrsim t$. The agreement deteriorates expectedly for smaller repulsions and temperatures, when magnetic ordering, spin correlations and charge fluctuations become more pronounced. In this case the used approximation gives the location of spectral maxima with comparatively large errors and fails in the description of fine structure of spectra. However, even in this case the approach describes correctly positions and widths of spectral bands. The results reproduce qualitatively correctly general shapes of spectra, the momentum distribution and their variation with changing repulsion. The imaginary part of the self-energy was shown to diverge as the Hubbard repulsion approaches the critical value from below with the value of exponent equal to 0.75. This quantity is half as much again the exponent found earlier for the semi-elliptical initial band.

\section*{Acknowledgements}
This work was supported by the research project IUT2-27, the European Regional Development Fund TK114 and the Estonian Scientific Foundation (grant ETF9371).

\end{document}